\documentclass[a4paper,11pt]{article}
\usepackage{jinstpub} 
\usepackage[subrefformat=parens]{subcaption}
\usepackage{siunitx}

\title{\boldmath Lifetime study of the ColdADC for the Deep Underground Neutrino Experiment}

\author[a, 1]{Wenjie Wu,\note{Current address: Institute of Modern Physics, Chinese Academy of Sciences, Lanzhou 730000, China.}}
\author[a, 2]{Benjamin Jargowsky,\note{Current address: Department of Physics, Boston University, 590 Commonwealth Avenue, Boston, MA 02215, U.S.A.}}
\author[a]{Yiwen Xiao,}
\author[a]{Alejandro Yankelevich,}
\author[a, 3]{Jianming Bian,\note{Corresponding author.}}
\author[b]{Cheng-Ju Lin,}
\author[b]{Tarun Prakash,}
\author[c]{and David Christian}
\affiliation[a]{Department of Physics and Astronomy, University of California, Irvine\\
Irvine, CA 92697, U.S.A.}
\affiliation[b]{Lawrence Berkeley National Laboratory\\
1 Cyclotron Road, Berkeley, CA 94720, U.S.A.}
\affiliation[c]{Fermi National Accelerator Laboratory\\
Batavia, IL 60510, U.S.A}

\emailAdd{bianjm@uci.edu}

\abstract{ColdADC is a custom ASIC digitizer implemented in 65 nm CMOS technology using specialized techniques for long-term reliability in cryogenic environments. ColdADC was developed for use in the DUNE Far Detector complex, which will consist of four liquid argon time projection chambers. Each contains 17 kilotons liquid argon as the target material in order to measure neutrino oscillations. Approximately 40,000 ColdADC ASICs will be installed for DUNE in the first two large detectors and will be operated at cryogenic temperatures during the experiment without replacement. The lifetime of the ColdADC is a critical parameter affecting the data quality and physics sensitivity of the experiment. A measurement of the lifetime of the ColdADC was carried out, and the results shown in this paper assure orders of magnitude longer lifetime of the ColdADC than the planned operation time of the detectors.}

\keywords{Neutrino detectors, Time projection chambers, Cryogenic detectors, Front-end electronics for detector readout}

\arxivnumber{2507.07086} 

\begin{document}
\maketitle
\flushbottom

\section{Introduction}
\label{sec:intro}
The Deep Underground Neutrino Experiment (DUNE) is a long-baseline accelerator-based neutrino experiment. The primary goals of DUNE are to definitively resolve the neutrino mass ordering puzzle and measure the CP violating phase \cite{duneTDR_physics}. These measurements will deepen our understanding of the symmetry in the neutrino sector and provide insight into the matter-antimatter asymmetry in the universe. DUNE will utilize four liquid argon time projection chambers (LArTPCs) as its Far Detector complex to search for neutrino interactions. The LArTPC technology enables excellent resolution for kiloton-scale particle detectors, which leads to high efficiency in differentiating signal events from background. It not only improves the sensitivity of neutrino oscillation measurements but also broadens the physics program.

The LArTPCs will be operated at 87 K to maintain the liquid argon in the liquid phase \cite{duneTDR_detector}. The electrical signal produced by the drift of the charges will be read out by a cold electronics system, which will be mounted directly on the LArTPC anodes and be immersed in LAr to achieve
optimal noise performance. Locating the cold electronics inside the cryostat also minimizes the amount of required cabling, as reading out $O(10^5)$ channels without any multiplexing within the detector would not be feasible and would cause unnecessary outgassing. The cold electronics system will consist of a front-end mother board (FEMB), equipped with three types of application-specific integrated circuits (ASICs) designed for best performance and lifetime in LAr. 
The first of these is a 16-channel ASIC for amplification and pulse shaping (referred to as LArASIC), designed with a 180 nm CMOS process \cite{larasic}. 
The second is the digitizer, called ColdADC \cite{coldadc_p2}.
The last is a 64-channel control and communications ASIC (referred to as COLDATA) \cite{coldata}. Both ColdADC and COLDATA are designed with a 65 nm CMOS process. Due to the different technology, LArASIC and ColdADC are designed separately instead of being integrated into a single ASIC.
For each FEMB, there will be eight LArASICs, eight ColdADCs, and two COLDATAs.

The ColdADC is a 16-channel, 16-bit, 2 MS/s digitizer. It receives input from one LArASIC. The input can be either single-ended or differential, and all inputs are sampled simultaneously. Two groups of 8 channels are multiplexed and input to two 15-stage pipelined ADCs. The ADC outputs are sent to COLDATA for further aggregation and transmission via copper links to warm electronics located outside the cryostat.
The current ColdADC design known as P2 is an improved version of the original P1 ASIC design, which was found to have several shortcomings through extensive testing, among which was an issue with the self-calibration. A detailed description of the design and the performance studies can be found in the literature \cite{coldadc_p2}. 
Approximately 40,000 ColdADCs will be installed into the DUNE Far Detector, and will be operated at cryogenic temperatures for several decades during DUNE's operation without replacement \cite{duneTDR_detector}. Therefore, the lifetime of the ColdADC is a critical parameter affecting the data quality and physics sensitivity. A thorough study of the lifetime of the ColdADC was carried out and described in this paper. The results assure orders of magnitude longer lifetime of the ColdADC than the planned operation time of the detectors.

\section{ColdADC lifetime study methodology}
\label{sec:method}
At cryogenic temperatures, many of the failure modes of CMOS devices become negligible, such as electromigration and stress migration. The remaining mechanism that may affect the lifetime of CMOS devices at low temperatures is the degradation induced by the hot carrier effect \cite{Kim_HCE, LArTPC_CMOS_Lifetime, nMOS_Lifetime}. Charge carriers in the channel of a MOS transistor are accelerated by the electric field resulting from the source-drain potential difference.  At cryogenic temperature, the distribution of kinetic energy of the charge carriers shifts to higher energy than at room temperature, and a larger fraction of the carriers acquire a large enough kinetic energy to ionize silicon.  The ionization electrons can escape the channel and become trapped in the gate oxide or at the interface between the gate oxide and the channel, causing changes in transistor characteristics that are similar to those caused by radiation damage.  This damage is called the hot carrier effect (HCE) and is especially important for nMOS transistors.

ColdADC was designed using a 65 nm CMOS process that has been shown to be relatively insensitive to HCE following extensive empirical studies \cite{BNL_SBND_ADC_Lifetime}. In response to these studies, two additional custom design rules were used to further mitigate the risk of damage due to HCE: the minimum transistor channel length was increased from 60 nm to 90 nm, and the bias voltages used were decreased from 2.5 V to 2.25 V for thick oxide transistors and from 1.2 V to 1.1 V for standard transistors.  Both of these design rules reduce the maximum electric field that can be present in the shortest transistors \cite{coldadc_p2}. A custom digital
standard cell library was also developed for the 1.1 V devices incorporating these rules.

This study of HCE-related damage was conducted to verify that ColdADC will not fail due to HCE during the lifetime of DUNE. However, the aging mechanism does not typically cause sudden device failure. Therefore, the lifetime due to HCE aging is defined by a chosen level of monotonic degradation of measured device parameters. The device is considered as failed if the chosen parameter falls outside of the specified range. In order to determine this monotonic degradation criterion for HCE, many efforts have been carried out. One criterion found to be effective is the current of the power supply voltage, which exhibits the desired monotonicity \cite{BNL_SBND_ADC_Lifetime}. Specifically, the current decreases more rapidly at higher stress voltages, allowing the lifetime measurements to be completed within a reasonable timeframe.

In order to minimize coupling between digital and analog circuitry, most of the ColdADC sub-circuits are laid out in deep n-wells. There are four isolated power domains in ColdADC as listed in table \ref{tab:coldadc_power}. All analog circuitry is powered by VDDA2P5 and is contained within a single large deep n-well. The digital circuitry responsible for directly governing analog functional blocks operates on VDDD2P5. The majority of other digital circuitry encompassing configuration registers, calibration and correction logic, clock generation, and similar components rely on power from VDDD1P2. The LVDS drivers, CMOS drivers, and receivers are energized by VDDIO. 

\begin{table}[t]
\centering
\caption{ColdADC power domains, adapted from table 1 in \cite{coldadc_p2}. \label{tab:coldadc_power}}
\smallskip
\begin{tabular}{c|c|c}
\hline
Power Domain Name & Description & Nominal value \\
\hline
VDDA2P5 & Analog Power & 2.25 V \\
VDDD2P5 & ADC Digital Power (Logic \& Switches) & 2.25 V \\
VDDD1P2 & Digital Logic Power & 1.1 V \\
VDDIO & ESD Ring/CMOS I/O Power & 2.25 V \\
\hline
\end{tabular}
\end{table}

The digital circuitry of ColdADC is much less likely to suffer HCE-related damage than the analog circuity since high fields are present in the channel of transistors in digital circuits only while they are switching.  Therefore, this study concentrated on the analog circuitry of ColdADC.  The chip being tested was stressed by repeatedly elevating the analog voltage supply (VDDA2P5) for a period of time; the analog voltage was then reduced to its normal value and a variety of performance measurements were made. Five ColdADC samples were tested at cryogenic temperature at four different voltages: 3.6 V, 3.8 V, 4.0 V, and two samples at 4.2 V. Following \cite{BNL_SBND_ADC_Lifetime}, ``failure'' was arbitrarily defined as the point where the current draw drops by 1\% so that a lower limit on the expected lifetime of ColdADC (due to HCE) can be extracted.

\begin{figure}[t]
\centering
\begin{subfigure}[t]{0.42\linewidth}
\centering	
\includegraphics[width=1.\textwidth]{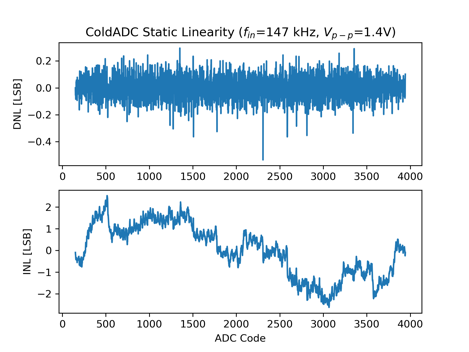}   
\caption{An example of DNL and INL.}\label{figinldnl}
\end{subfigure}
\begin{subfigure}[t]{0.42\linewidth}
\centering	
\includegraphics[width=1.\textwidth]{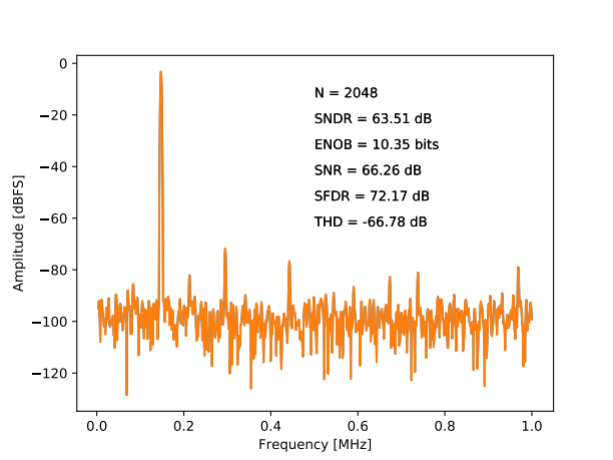}
\caption{An example of the FFT results from the measured sine wave.}\label{figfft}
\end{subfigure}
\caption{Examples of the observation parameters.}
\end{figure}

Several performance metrics were periodically evaluated. The differential non-linearity (DNL) and integral non-linearity (INL) are the metrics for the static linearity of the ColdADC, which were measured with the simulated input from the function generator. The measured histogram of ADC codes was fitted with a sine function. The residuals from the fit yields the DNL as a function of ADC code, and the integral of DNL is the INL. An example of these two distributions is shown in figure \ref{figinldnl}. For these plots, the ColdADC output was truncated to 12 bits.

A fast Fourier transform (FFT) of the response to the input sine wave was also performed (figure \ref{figfft}), and standard ADC figures of merit were extracted. These metrics include the signal to noise and distortion ratio (SNDR),  spurious free dynamic range (SFDR),  the total harmonic distortion (THD), and the equivalent number of bits (ENOB). ENOB is the ratio of the input signal amplitude to the RMS sum of all other spectral components. It represents a comprehensive measure of the combination of nonlinearity and noise and is therefore of particular importance.

\section{ColdADC lifetime test setup}
\label{sec:setup}
\begin{figure*}[t]
\centering
	\includegraphics[width=0.37\textwidth]{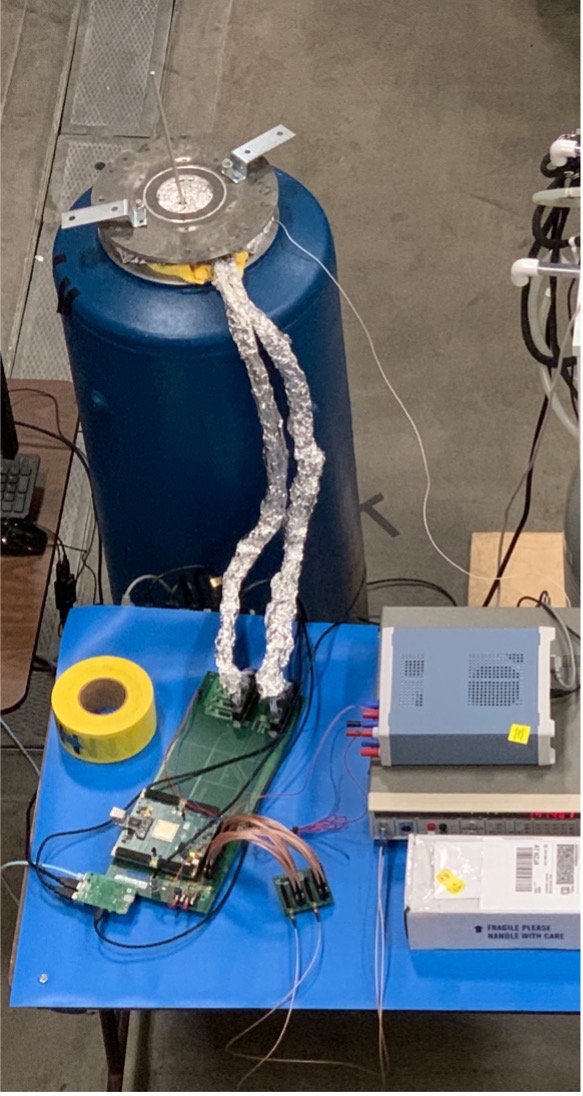}
	\caption{Overhead view of the testing setup.}
    \label{overview}
\end{figure*}

\begin{figure}
\begin{subfigure}[t]{0.5\linewidth}
\centering	
\includegraphics[width=1.\textwidth]{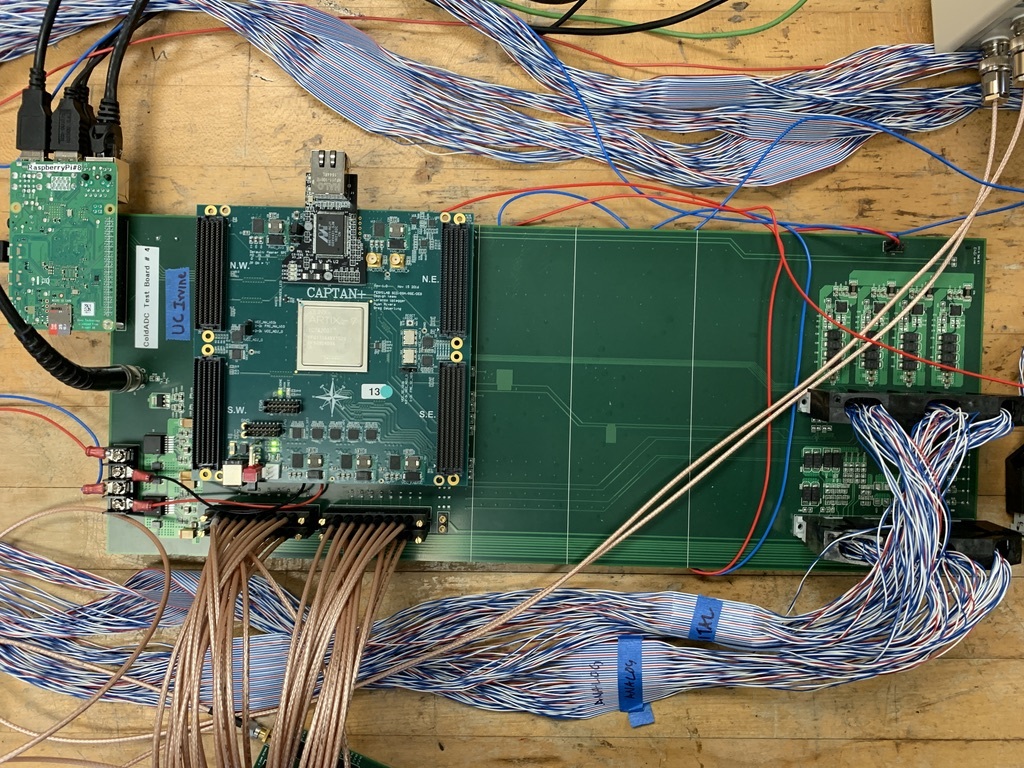}
\caption{Motherboard with peripherals attached, not in use in this photo.}\label{figmotherboard}
\end{subfigure}
\begin{subfigure}[t]{0.5\linewidth}
\centering	
\includegraphics[width=1.\textwidth]{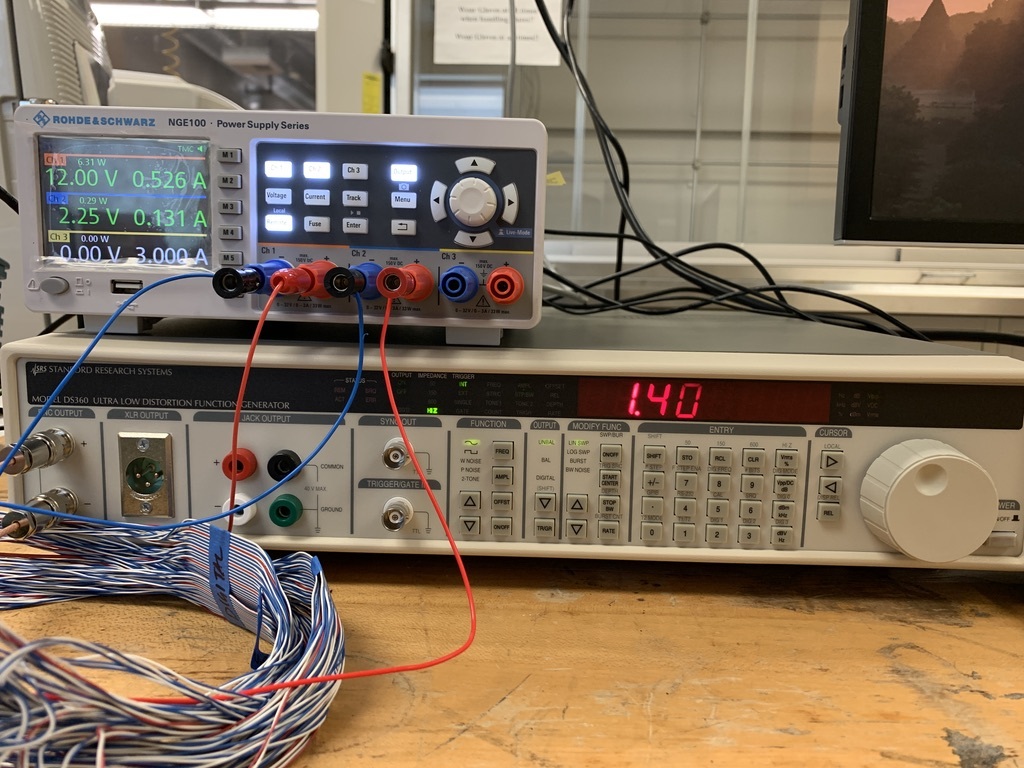}
\caption{The power supply on top of the function generator.}\label{figgenerator}
\end{subfigure}
\begin{subfigure}[t]{0.5\linewidth}
\centering	
\includegraphics[width=1.\textwidth]{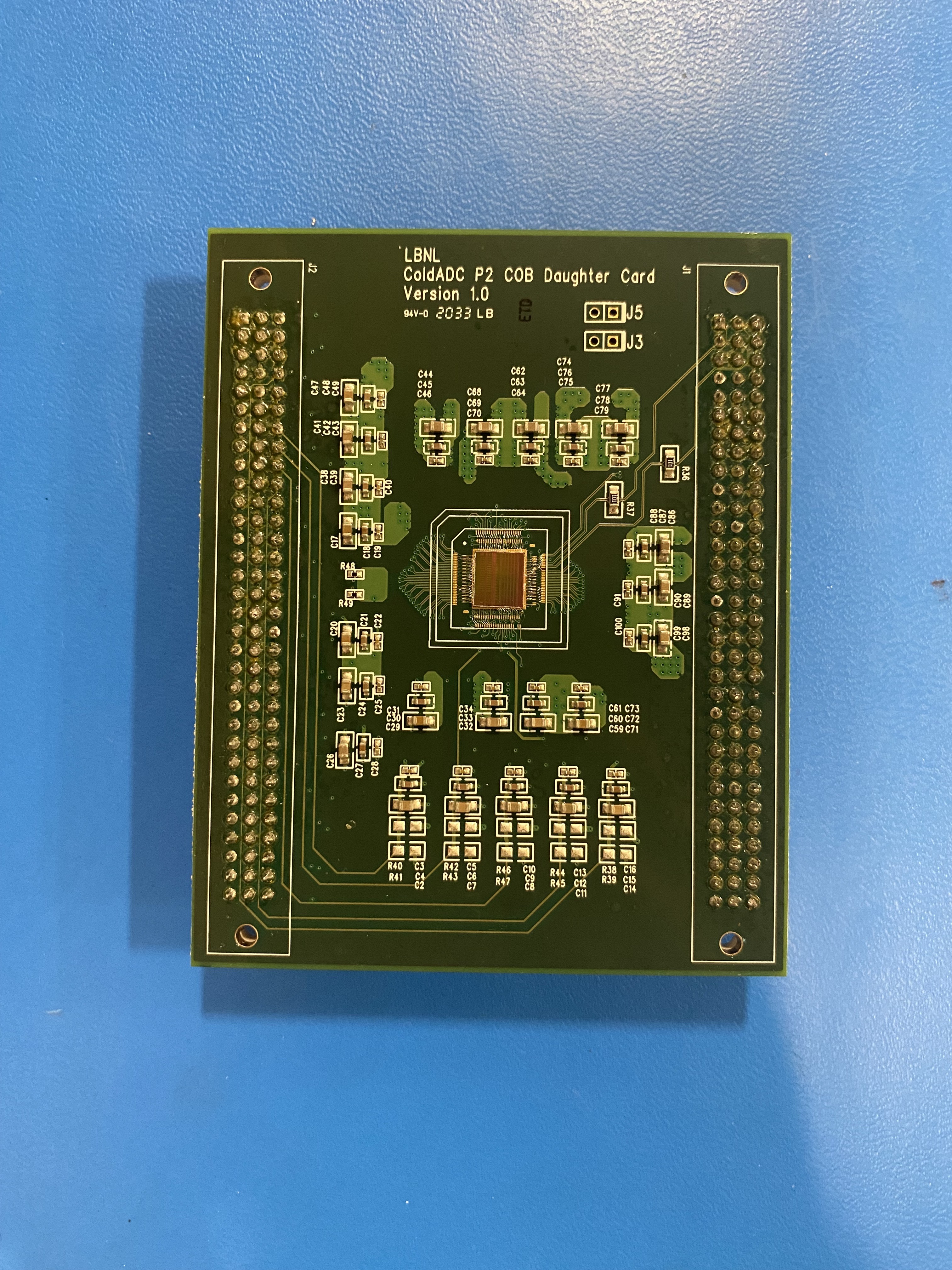}
\caption{Daughter card with ColdADC in the center.}\label{figdaughtercard}
\end{subfigure}
\begin{subfigure}[t]{0.5\linewidth}
\centering	
\includegraphics[width=1.\textwidth]{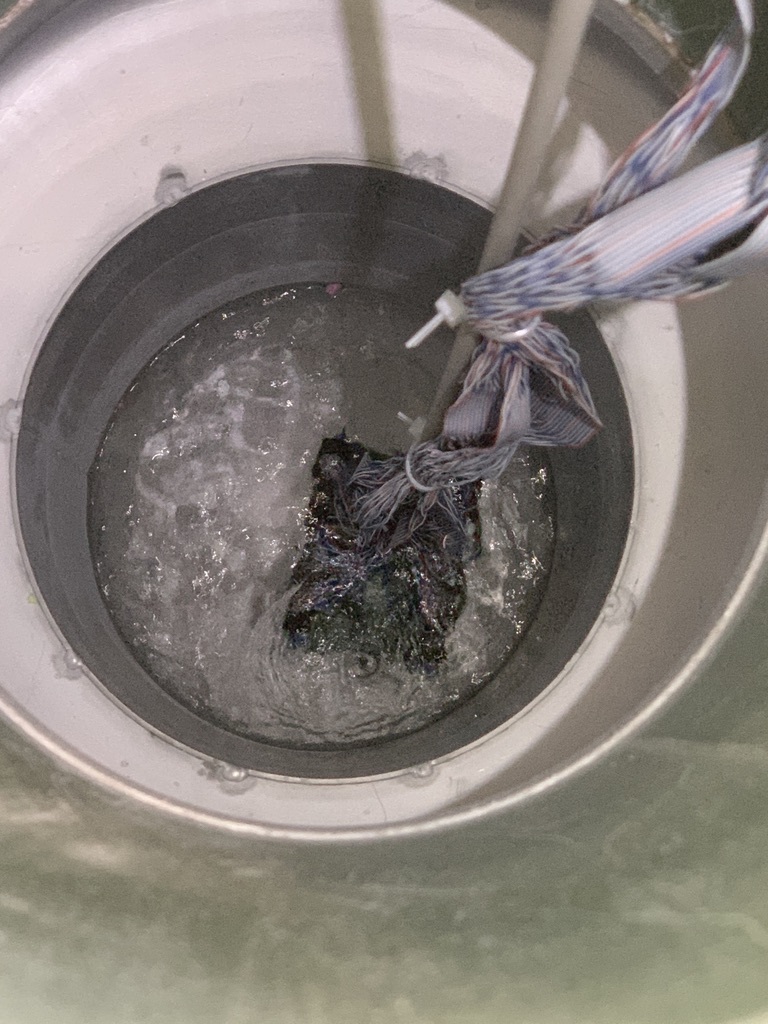}
\caption{Daughter card submerged in liquid nitrogen in dewar, viewed from top. The top of the dewar is not covered here for the picture as it usually was during the experiment.}\label{figdaugtercardimmersed}
\end{subfigure}
\caption{Components of the test setup.}
\end{figure}

The ColdADC lifetime study test setup is shown in figure \ref{overview}. A 50 L dewar, shown at the top of the picture, stores $\mathrm{LN}_2$ for the test. $\mathrm{LN}_2$ is used for cost reduction purposes since the 77 K temperature of $\mathrm{LN}_2$ is comparable to the 87 K temperature of LAr. The dewar's large capacity 
extends the $\mathrm{LN}_2$ refilling cycle, reducing interruptions from non-relevant activities. A custom motherboard shown in figure \ref{figmotherboard} is developed for the lifetime test. The ColdADC is mounted on a daughter card, as shown in figure \ref{figdaughtercard}, which is immersed in $\mathrm{LN}_2$ within the dewar, as shown in figure \ref{figdaugtercardimmersed}. The analog and digital signals of the ColdADC are transmitted through the two long cables that connect the daughter card and the motherboard. Both cables are wrapped in aluminum foil to minimize the pickup noise from the environment. A CAPTAN + FPGA mezzanine board \cite{CAPTAN} is used to collect and transfer data. The system is controlled by a commercial Raspberry Pi board attached to the motherboard through its GPIO pinouts. The test components are placed on an electrostatic discharge (ESD) mat, which is grounded to the AC outlet. All manual operations are performed with a wrist strap connected to the mat in order to reduce static damage to the test components.

\begin{figure*}[t]
\centering
	\includegraphics[width=1.0\textwidth]{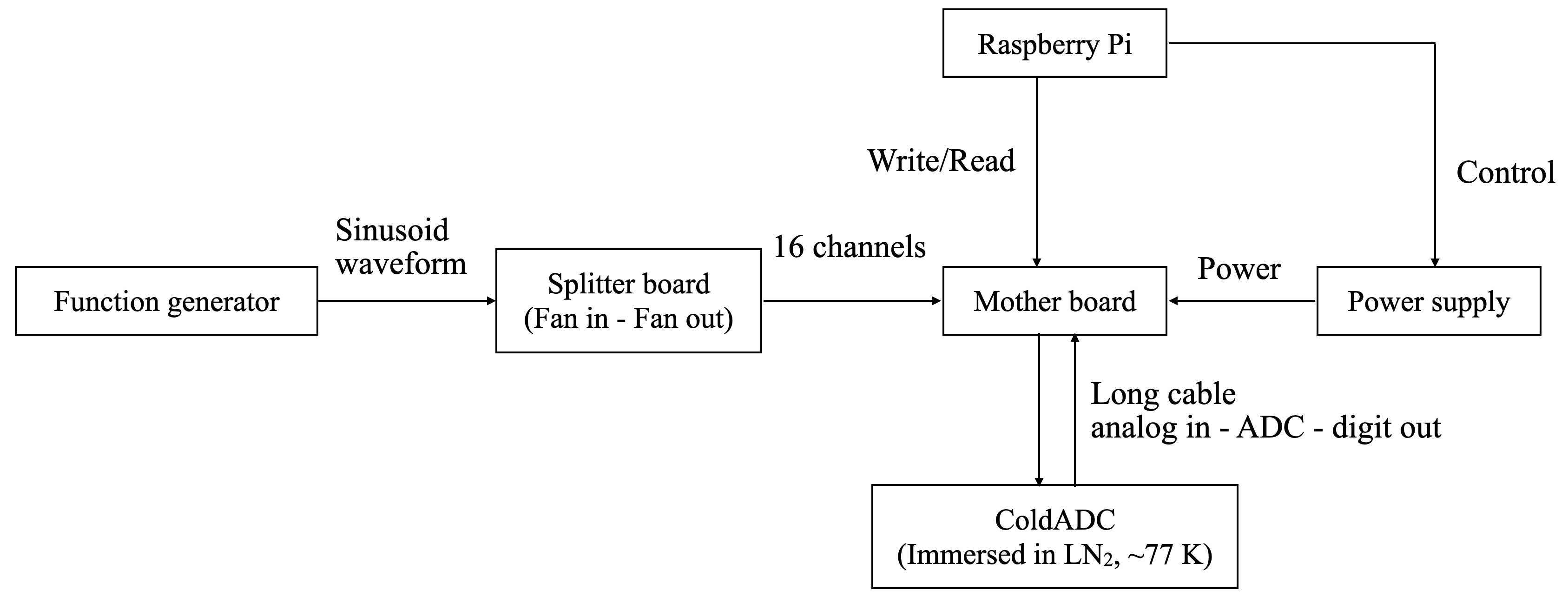}
	\caption{Schematic diagram of the test setup.}
    \label{schematic}
\end{figure*}

A schematic diagram of the test setup is shown in figure \ref{schematic}. A sine waveform (147.461 kHz, 1.4 Vpp, 0.9 Vdc offset) is generated by a low distortion function generator (Stanford Research DS360), as shown in figure \ref{figgenerator}. This waveform is used as the input signal to test the non-linearity of the ColdADC. The frequency is determined for the purpose of coherent sampling, and the voltage range is determined by the dynamic range of ColdADC. The signal is split into 16 channels to test all channels simultaneously. The motherboard controls the streams of input and output signals and communicates with the Raspberry Pi where the readout data is stored. The motherboard is powered by an external power supply (The Rohde \& Schwarz NGE 100), which also provides the bias voltage to VDDA2P5. The power supply can measure the current flow through VDDA2P5 and transmit data via an USB interface. An example of a current measurement is shown in figure \ref{figgenerator}. A python-based slow control interface runs on the Raspberry Pi for automatic data taking and analysis to minimize human intervention during the test.

\begin{figure}
\begin{subfigure}[t]{1.\linewidth}
\centering	
\includegraphics[width=1.0\textwidth]{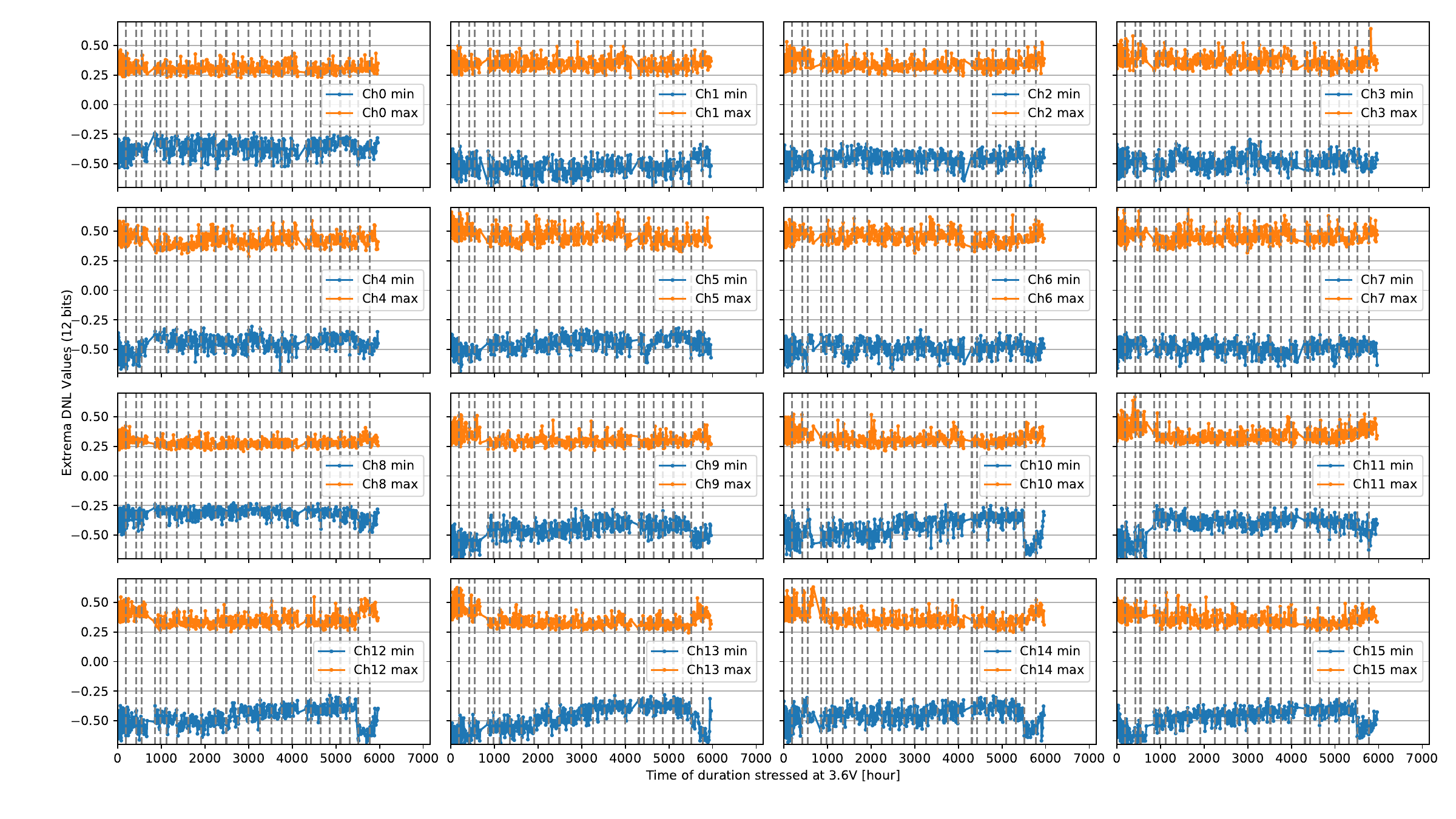}
\caption{Variation of the extrema of DNL.}\label{figmindnl}
\end{subfigure}
\begin{subfigure}[t]{1.\linewidth}
\centering	
\includegraphics[width=1.0\textwidth]{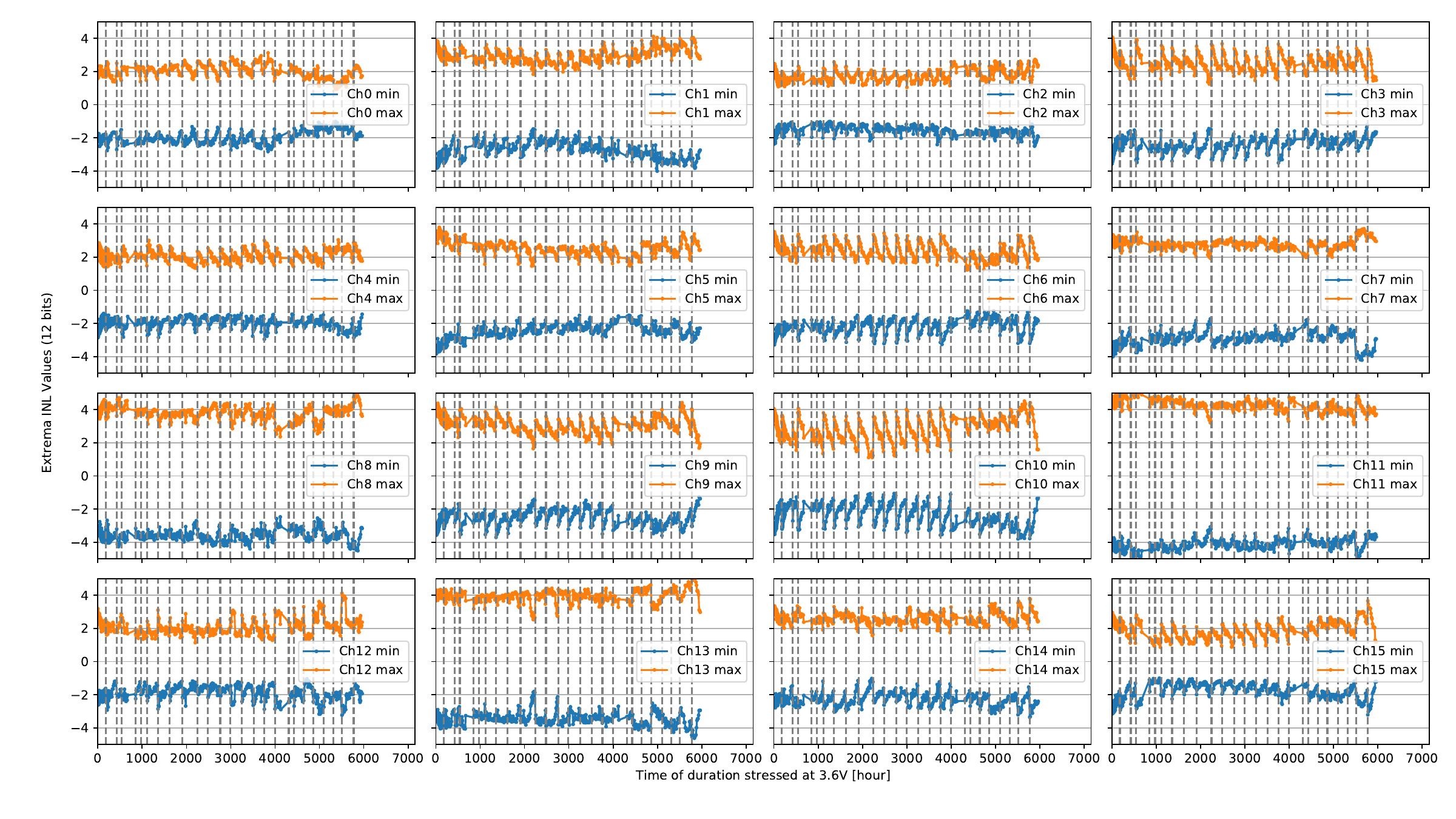}
\caption{Variation of the extrema of INL.}\label{figmininl}
\end{subfigure}
\caption{Maximum and minimum DNL (a) and INL (b) values varying in time, per channel. The dashed vertical lines indicate times when the dewar was refilled.}\label{figsinldnl}
\end{figure}

\begin{figure}[htbp]
\centering
\includegraphics[width=0.8\textwidth]{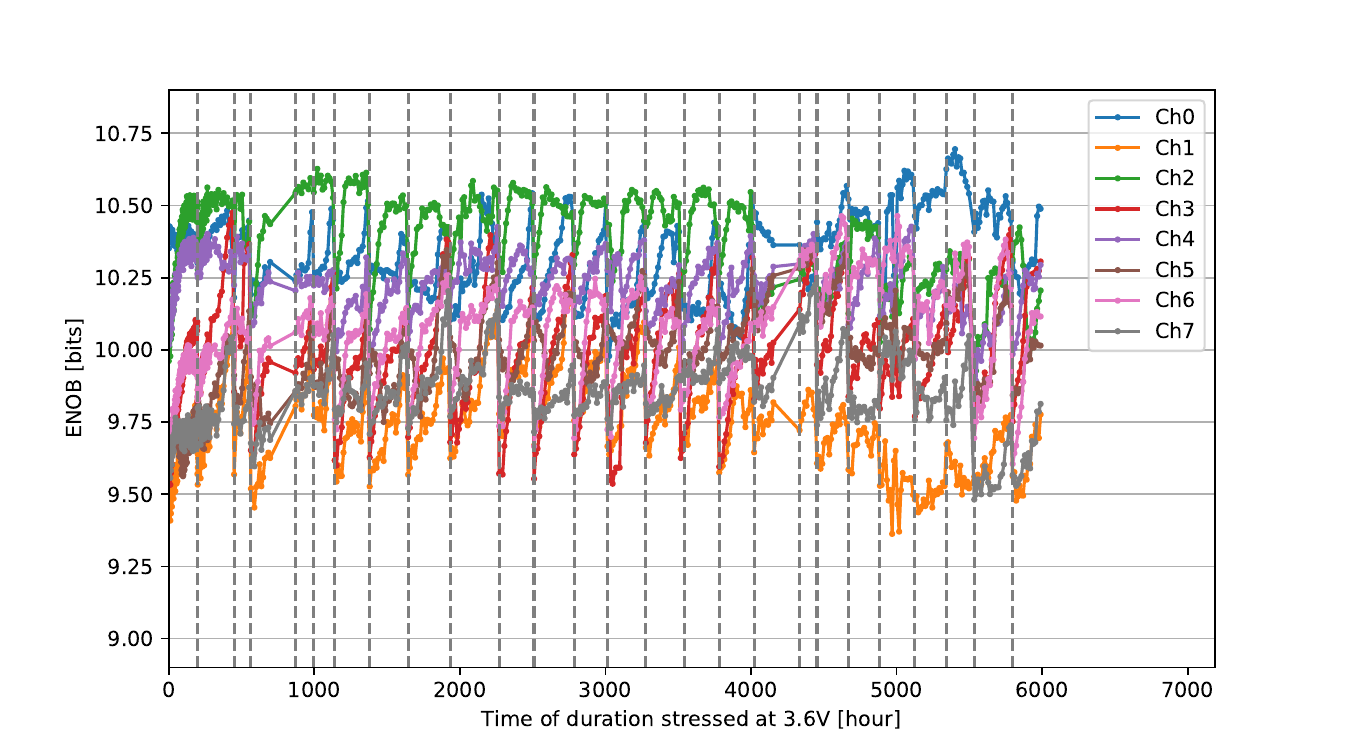}
\includegraphics[width=0.8\textwidth]{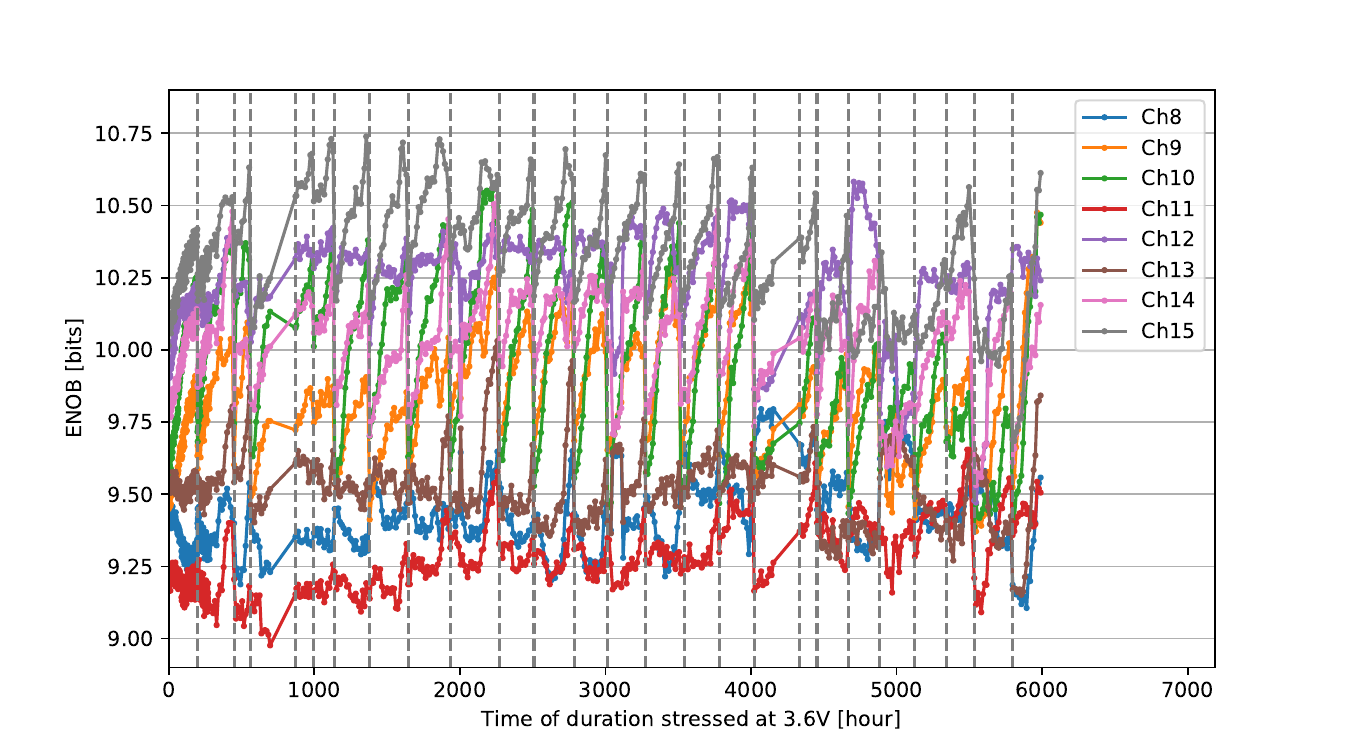}
\caption{ENOB varying over time for each of the 16 channels, broken up by first and last 8 channels.}
\label{ENOB_time}
\end{figure}

As outlined in section \ref{sec:method}, DNL, INL, and ENOB measured at the nominal voltage are recorded throughout the test for each test ColdADC. The variations of these parameters over time for the ColdADC sample stressed at 3.6 V  
are shown in figures \ref{figsinldnl} and \ref{ENOB_time}. This ColdADC sample was stressed over 6000 hours at the cryogenic temperature (77 K). Every 12 hours, the applied voltage was momentarily reduced to the nominal 2.25 V for 5 minutes. During this time, the ADC values are recalibrated, and function generator data is collected as described in section \ref{sec:method}.

\begin{figure}[t]
\centering
\includegraphics[width=.8\textwidth]{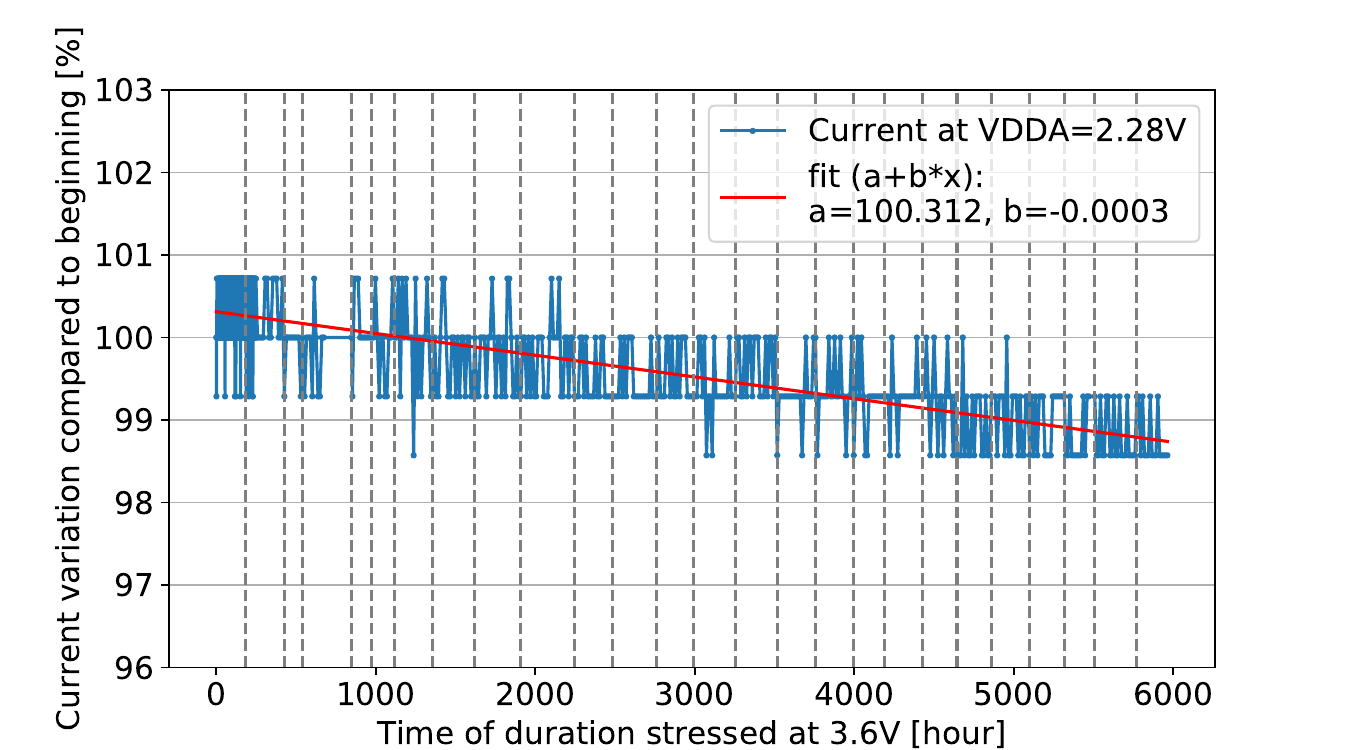}
\caption{Current variation compared to the beginning.}	
\label{fig_currentvariation_current}
\end{figure}

Normally the dewar needs to be refilled with $\mathrm{LN}_2$ about every 10 days, and the dashed vertical lines in figures \ref{figsinldnl}--\ref{fig_currentvariation_current} indicate the time of refilling. There were two incidents affecting the data taking for this ColdADC sample. The first was a system crash of the Raspberry Pi that prevented the stress voltage from being applied from 2/25/2022 to 3/1/2022 (89.5 hours in total). The second was due to a shortage of $\mathrm{LN}_2$ in the lab. The system was stopped manually during this shortage for 23 hours, from 5/1/2022 to 5/2/2022. The correction of the stressed time due to data taking and the incidents was applied in these figures.

\section{Results and conclusions}
Figure \ref{figmindnl} shows the variation of the extrema of DNL (blue for the minima and orange for the maxima) of all ADC codes for each channel, and figure \ref{figmininl} shows the variation of the extrema of INL. After every refill, there is a sudden change due to the location shift of the ColdADC, the cables, and the shielding wrap. These shifts cause slight changes in the environmental pickup. However, the overall change is within a small range, which indicates the ColdADC is functioning correctly and has consistent performance over time. The ENOBs of all 16 channels shown in figure \ref{ENOB_time} have the same pattern. Most of the ENOBs vary within 0.5 bits, which implies there is no degradation of the performance in terms of the non-linearity and noise.

\begin{figure}[htbp]
\centering
\includegraphics[width=0.6\textwidth]{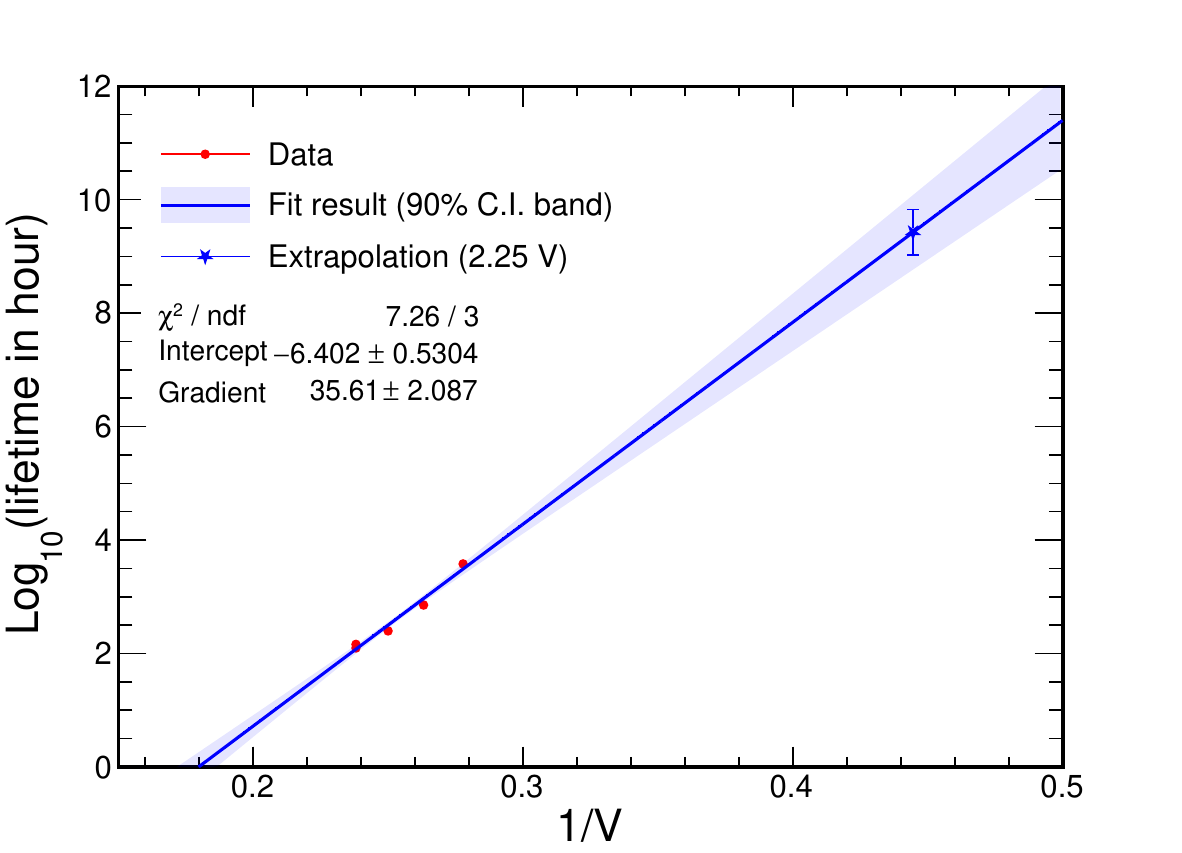}
\caption{Lifetime as a function of the inverse of the stress voltage. The projection of the lifetime at 2.25 V is done by extrapolating the fitting results. This result represents a conservative lower bound of the lifetime.}
\label{fig_lifetime_extrapolation}
\end{figure}

The current draw shown in figure \ref{fig_currentvariation_current} shows a general decrease over the course of the test, and we fit the results assuming that the current falls linearly with time. The slope of the fitting results is the percentage of current drop per hour. The fitting results of all five ColdADC samples are shown in table \ref{tab:fittingResults}. The uncertainties are obtained from the fit and are assumed to be uncorrelated. As explained in section \ref{sec:method}, the degradation threshold was determined as a 1\% drop in the current draw. In the absence of any other measurable degradation in performance, this threshold represents a conservative lower bound on the lifetime. The difference between two measurements stressed at 4.2 V is considered as the uncertainty arising from sample variation. The relationship between the lifetime and the inverse of the stress voltage can be described by an exponential function \cite{Kim_HCE}, as is evident in figure \ref{fig_lifetime_extrapolation}. A linear fit assuming $\log_{10} \tau\propto 1/V$ is performed, and this fitting result is then extrapolated to different 1/V values. This approach gives a lifetime of $3.02\times10^5$ years at the nominal ColdADC operation voltage, which is orders of magnitude longer than DUNE’s lifetime.

\begin{table}[h]
\centering
\caption{Results of linear fits ($a+bx$) to the current variation vs time for the five ColdADC samples.\label{tab:fittingResults}}
\smallskip
\begin{tabular}{c|c|c}
\hline
Stress voltage [V] & Fitting result of the slope (b) & Fitting uncertainty of the slope \\
\hline
3.6 & $2.6\times10^{-4}$ & $7.8\times10^{-6}$ \\
3.8 & $1.4\times10^{-3}$ & $9.3\times10^{-5}$ \\
4.0 & $4.0\times10^{-3}$ & $2.8\times10^{-4}$ \\
4.2 & $8.0\times10^{-3}$ & $3.2\times10^{-4}$ \\
4.2 & $6.9\times10^{-3}$ & $3.8\times10^{-4}$ \\
\hline
\end{tabular}
\end{table}

\newpage

\acknowledgments

This document was prepared by members of the DUNE collaboration using the resources of the Fermi National Accelerator Laboratory (Fermilab), a U.S. Department of Energy, Office of Science, Office of High Energy Physics HEP User Facility. Fermilab is managed by FermiForward Discovery Group, LLC, acting under Contract No. 89243024CSC000002.



\bibliographystyle{JHEP}

\end{document}